\documentclass[conference,letter,10pt]{IEEEtran}

\usepackage{amsmath, amssymb, amscd, amsthm, amsfonts}
\usepackage[pdftex]{graphicx}
\usepackage[caption=false,font=footnotesize]{subfig}
\usepackage{dblfloatfix}    
\usepackage{color}
\usepackage{algorithm}
\usepackage[noend]{algpseudocode}
\usepackage{comment}
\usepackage{hyperref}
\usepackage{tikz,ifthen}
\usetikzlibrary{positioning,arrows,backgrounds}
\usetikzlibrary{fit}
\usetikzlibrary{calc}

\newtheorem{theorem}{Theorem}
\newtheorem{lemma}{Lemma}

\newtheorem{proposition}{Proposition}

\usepackage{pgfplots}
\pgfplotsset{width=10cm,compat=1.9}

\usepackage{subfiles}
\usepackage[noadjust]{cite}

\usepackage[english]{babel}

\addto\captionsenglish{}

\graphicspath{{../Figures/}}



\begin{document}

\title{Is Phase Shift Keying Optimal for Channels with Phase-Quantized Output?}

\author{
   \IEEEauthorblockN{Neil Irwin Bernardo, Jingge Zhu, and Jamie Evans
   }
   \IEEEauthorblockA{Department of Electrical and Electronics Engineering, University of Melbourne\\
   }
   
   \IEEEauthorblockA{
      Email: bernardon@student.unimelb.edu.au,\{ jingge.zhu, jse\}@unimelb.edu.au
   }
}
\maketitle
\begin{abstract}
This paper establishes the capacity of additive white Gaussian noise (AWGN) channels with phase-quantized output. We show that a rotated $2^b$-phase shift keying scheme is the capacity-achieving input distribution for a complex AWGN channel with $b$-bit phase quantization. The result is then used to establish the expression for the channel capacity as a function of average power constraint $P$ and quantization bits $b$. The outage performance of phase-quantized system is also investigated for the case of Rayleigh fading when the channel state information (CSI) is only known at the receiver. Our findings suggest the existence of a threshold in the rate $R$, above which the outage exponent of the outage probability changes abruptly. In fact, this threshold effect in the outage exponent causes $2^b$-PSK to have suboptimal outage performance at high SNR.
\end{abstract}


\IEEEpeerreviewmaketitle

\section{Introduction}\label{section-intro}

The use of low-resolution analog-to-digital converters (ADCs) has recently gained significant research interest because it addresses practical problems and scalability issues in 5G core technologies such as massive data processing, high power consumption, and cost \cite{Liu:2019}. Most studies on low-resolution ADCs have been more focused on investigating the fundamental limits and practical detection strategies in the context of multiple-input multiple-output (MIMO) and millimeter wave (mmWave) systems \cite{Jacobsson:2015,Bjornson:2015,Orhan:2015,Mezghani:2012}. However, these studies did not properly address the structure of the capacity-achieving input and only analyzed performance via capacity bounds using simplified analytical models. Low-resolution receiver design requires a shift in signal/code construction since Gaussian signaling is no longer optimal in channels with quantized output \cite{Vu:2019}.

Some research efforts have been invested in analyzing the capacity limits of channels with low-resolution quantization and finding the optimum signaling schemes for such channels. One of the first studies on this topic showed that binary antipodal signaling is optimal for real AWGN channel with 1-bit quantized output \cite{Singh:2009}. Extension of capacity analysis to other wireless channels with 1-bit in-phase and quadrature (I/Q) output quantization revealed that QPSK is the optimum signaling for coherent/noncoherent Rayleigh channel \cite{Mezghani2:2008}\cite{Krone:2010}, noncoherent Rician channel \cite{Vu:2019}, and zero-mean Gaussian mixture channel \cite{Rahman:2020}. However, identifying the structure of capacity-achieving input analytically for static and fading channels with multi-bit I/Q quantization still remains an open problem \cite{Vu:2019}.  

Motivated by the above discussion, we aim to extend the capacity results of 1-bit I/Q quantization to multi-bit quantization. However, we shall investigate multi-bit phase quantization instead of the conventional I/Q quantization. Phase quantization ignores the amplitude component thus eliminating the necessity for automatic gain control \cite{Singh:2013}. Furthermore, phase quantizers can be easily implemented in practice using analog phase detectors and 1-bit comparators which consume negligible power (in the order of mW) \cite{Gayan:2020}. Information rate of phase-quantized block noncoherent receiver has been studied before in \cite{Singh:2013} but the proponents of the study did not show the optimality of phase shift keying. Error rate analysis of low-resolution phase-modulated communication has been done for the single-input single-output (SISO) fading channel \cite{Gayan:2019,Gayan:2020}, relay channel \cite{Souryal:2008}, and multiuser MIMO channel \cite{Lopes:2020} but only investigated uncoded transmissions. In this work, we provide a rigourous proof that phase shift keying is indeed capacity-achieving for static channels with phase-quantized output. In fact, the analytical tractability of the 1-bit ADC case in \cite{Mezghani2:2008,Singh:2009,Vu:2019,Krone:2010,Rahman:2020} comes from the tractability of the more general multi-bit phase quantization. We then extend the analysis to phase-quantized Rayleigh fading channel with channel state information (CSI) known only at the receiver and give some insights about the outage exponent (or diversity order) of its outage probability. In particular, our numerical results reveal a threshold effect in the outage exponent when the required transmission rate $R$ of an $M$-PSK scheme exceeds a certain value. The proofs can be found in the supplement material \cite{ISIT2021_supplement}.

\section{System Model}\label{section-sys_model}

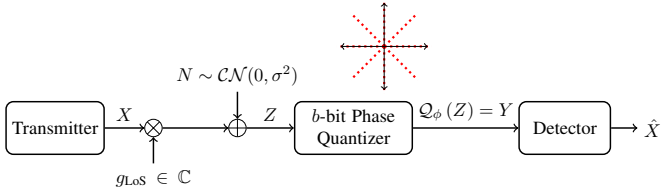
\begin{figure}[t]
    \centering
    \scalebox{.7}{
    \begin{tikzpicture}
     \tikzstyle{multiply_block} = [circle,{path picture={ 
      \draw[black]
    (path picture bounding box.south east) -- (path picture bounding box.north west) (path picture bounding box.south west) -- (path picture bounding box.north east);
    }}]
    
    \tikzstyle{add_block} = [circle,{path picture={ 
      \draw[black]
    (path picture bounding box.south) -- (path picture bounding box.north) (path picture bounding box.west) -- (path picture bounding box.east);
    }}]
    
    \tikzset{
        pre/.style =    {<-, semithick},
        post/.style =   {->, semithick}
    }

    \tikzset{block/.style = {draw, fill=white, rectangle,
                  minimum height=3em, minimum width=2cm,text width=1.5cm,align=center},
        input/.style = {coordinate},
        output/.style = {coordinate},
        pinstyle/.style = {pin edge={to-,t,black}}
    }

    \node[draw,rectangle,rounded corners,minimum height = 1cm, minimum width = .75cm,thick] (Tx)      {Transmitter};
    \node[draw,multiply_block] (multiplier)    [right =.75cm of Tx] {};
    \node[draw,add_block] (adder)         [right= 1.25cm of multiplier]   {} edge[pre,thick] (multiplier);
    \draw[->,thick] (Tx.east) -- (multiplier.west) node[pos = .5,above] {$X$};


    \node[] (noise)         [above of=adder]        {$N \sim \mathcal{CN}(0,\sigma^2)$} edge[post,thick] (adder);

      \node[draw,rectangle,rounded corners,minimum height = 1cm, minimum width = .75cm] (ADC)           [right=.9cm of adder,text width = 2cm,align=center,thick]        {$b$-bit Phase Quantizer} edge[pre,thick] node[auto,swap] {$Z$} (adder);
      
     \node[draw,rectangle,rounded corners,minimum height = 1cm, minimum width = .75cm,thick] (baseband_proc)           [right=2cm of ADC,text width = 1.5cm,align=center]        {Detector} edge[pre,thick] node[auto,swap] {$\mathcal{Q}_{\phi}\left(Z\right) = Y$} (ADC);
     
     \node[right=.5cm of baseband_proc] (Rx) {$\hat{X}$} edge[pre,thick] node[auto,swap] {} (baseband_proc);
     
     \node[below = .5cm of multiplier,text width = 2.5cm,align=center] {$g_{\text{LoS}}\in\mathbb{C}$} edge[->,thick] (multiplier.south);
     
    \begin{scope}[xshift = 6.25 cm,yshift = 1.55cm, scale = 0.15]
    
        \foreach \p in {0,45,...,359} {
   
        \draw[dotted,very thick, red,rotate = 0] (0,0) -- (\p:6);
          }
      \foreach \a in {0,90,...,359}
        \draw[thin, ->] (0,0) -- (\a:5.5);

    \end{scope}

\end{tikzpicture}
}
    \caption{System Model of Phase-Quantized Receiver}
    \label{fig:psk_receiver_sys_model}
\end{figure}

We consider a discrete-time baseband model shown in Figure \ref{fig:psk_receiver_sys_model}. The transmitter sends a signal $X$ which has an average power constraint $\mathbb{E}[|X|^2] \leq P$. $g_{\text{LoS}}$ is a complex constant representing the gain and direction of the line-of-sight (LoS) component and $N\sim \mathcal{CN}(0,\sigma^2)$ is an additive noise. We can express the received signal prior to quantization as
\begin{equation}
    Z = g_{\text{Los}}X + N.
\end{equation}
The signal is then fed to a symmetric $b$-bit phase quantizer $\mathcal{Q}_\phi(\cdot)$ to produce an integer-valued output $Y \in [0,2^b - 1]$. To be more precise, the output of the phase quantizer is $Y=y$ if $\angle Z \in \mathcal{R}^{\text{PH}}_{y}$, where $\mathcal{R}^{\text{PH}}_{y}$ is given by
\begin{equation*}
    \mathcal{R}^{\text{PH}}_{y} = \left\{\phi\in [-\pi,\pi]\;\Big|\;\frac{2\pi}{2^{b}}y - \pi \leq \phi < \frac{2\pi}{2^{b}}(y+1) -\pi \right\}.
\end{equation*}
Due to the circular structure of the phase quantizer, the addition operation $Y+k$ for some $k\in\mathbb{Z}$ constitutes a modulo $2^b$ addition. In this quantization model, only a coarse phase information of the received signal is retained and the goal of the receiver is to reliably recover the message encoded in $X$ using the phase quantizer output, $Y$. It should be noted that the discrete-time channel model we considered implicitly assumes that the phase quantizer is symmetric and that the channel output is sampled at the Nyquist rate. However, such quantization and sampling strategy may not be optimal in some cases as pointed out in \cite{Koch:2013,Koch:2010}.


We identify the probability quantities essential to express the mutual information. Suppose we define $U = \frac{g_{\text{LoS}}}{\sigma}X$ and $Z' = \frac{Z}{\sigma}$. The conditional PDF $p_{Z'|U}(z'|u)$ is given by
\begin{equation}
        p_{Z'|U}(z'|u) = \frac{1}{\pi }\exp\left(-|z'-u|^2\right).
\end{equation}
Note that in the phase-quantized receiver, we discard any information on the magnitude. Suppose we represent the random variables in polar form (i.e. use $Z' = \frac{\sqrt{R}}{\sigma}e^{j\Phi}$ and $U = \sqrt{\alpha} e^{j\Theta}$). The probability of $\angle Z' = \Phi$ given $U = \sqrt{\alpha}$ is transmitted can be written as
\begin{equation}\label{eq:prob_phi_given_u}
    \begin{split}
        p_{\Phi|A}(\phi|\alpha) =& \int_{R} p_{Z'|U}\left(z'=\frac{\sqrt{r}}{\sigma}e^{j\phi}\Big|u = \sqrt{\alpha}\right)\;dr\\
        =& \frac{e^{-\alpha}}{2\pi}+\frac{\sqrt{\alpha}\cos\phi e^{-\alpha\sin^2\phi}\left[1-Q\left(\sqrt{2\alpha}\cos\phi\right)\right]}{\sqrt{\pi}},\\
    \end{split}
\end{equation}
where the last equality is obtained from \cite[equation (10)]{Fu:2008}. $Q(\cdot)$ is the tail probability of the standard normal distribution. The conditional PDF $p_{Y|U}(y|u)$ (or $p_{Y|A,\Theta}(y|\alpha,\theta)$), denoted as $W_{y}^{(b)}(u)$ (or $W_{y}^{(b)}(\alpha,\theta)$), is given by \begin{align}\label{eq:p_y_given_u}
             W_{y}^{(b)}(\alpha,\theta) \; 
             =& \int_{\mathcal{R}^{\text{PH}}_{y}-\theta}p_{\Phi|A}(\phi|\alpha)\;d\phi \nonumber\\
             =&\int_{\frac{2\pi}{2^{b}}y - \pi -\theta}^{\frac{2\pi}{2^{b}}(y+1) -\pi -\theta }\;p_{\Phi|A}(\phi|\alpha)\;d\phi.
\end{align}
Equation (\ref{eq:p_y_given_u}) has no closed-form expression. However, we can still use it to identify the optimal input distribution and numerically compute the capacity of the phase-quantized system. Now, consider an input distribution $F_U(u)$ with density function $f_U(u)$. With slight abuse of notation, we use $F_U$ and $f_U$ to denote $F_U(u)$ and $f_U(u)$, respectively. For a given $F_U$, the probability mass function (PMF) of $Y$ is therefore
\begin{equation}\label{eq:pmf_y}
        \begin{split}
             p(y;F_U) =& \int_{\mathbb{C}} W_{y}^{(b)}(u)\;dF_U\;\;\forall y.
        \end{split}
\end{equation}
We use the above notation to emphasize that the PMF of $Y$ is induced by the choice of the distribution $F_U$. Given the above probability quantities, we can now express the mutual information between $U$ and $Y$ as follows:
\begin{align}\label{eq:MI_U_Y}
        I(U;Y) =& I(F_U) = H\left(Y\right) - H\left(Y|U\right),\\
        \text{where }\;\;H(Y)=& -\int_{\mathbb{C}}\sum_{y=0}^{2^b-1}W_{y}^{(b)}(u)\log p(y;F_U)\;dF_U\nonumber\\
        H(Y|U)=& -\int_{\mathbb{C}}\sum_{y=0}^{2^b-1}W_{y}^{(b)}(u)\log W_{y}^{(b)}(u)\;dF_U.\nonumber
\end{align}
We use the notation $I(F_U)$ since the mutual information is a result of choosing a specific input distribution $F_U$. All $\log(\cdot)$ functions in this paper are in base 2 unless stated otherwise. Let $P'=\frac{|g_{\text{LoS}}|^2}{\sigma^2}P$. The capacity for a given power constraint is the supremum of mutual information between $U$ and $Y$ over the set of all input distributions $F_U$ satisfying the power constraint $\mathbb{E}[|U|^2] \leq P'$. In other words,
\begin{equation}\label{eq:cap_def}
    C = \sup_{F_U \in \Omega} I(F_U) = I(F_U^*),
\end{equation}  
where $\Omega$ is the set of all input distributions which have average power less than or equal to $P'$. In the next section, we establish the properties of the capacity-achieving input distribution $F^*_U$ when $g_{\text{LoS}}$ is known at the transmitter.

\section{Capacity-achieving Input for Phase-Quantized AWGN Channel}\label{section-input_distribution}

The mutual information $I(F_U)$ is concave with respect to $F_U$ \cite[Theorem 2.7.4]{Cover:2006_IT} and the power constraint ensures that $\Omega$ is convex and compact with respect to weak* topology\footnote{This is the coarsest topology in which all linear functionals of $dF_U$ of the form $\int f(u)dF_U$, where $f(u)$ is a continuous function, are continuous.} \cite{Abou-Faycal:2001}. The existence of $F_U^*$ is equivalent to showing that $I(F_U)$ is continuous over $F_U$. The finite cardinality of phase quantizer output trivially ensures this and the proof follows closely to the method in \cite[Appendix A]{Singh:2009} and \cite[Lemma 1]{Vu:2019}. We first show that the optimal input distribution satisfies a certain phase symmetry. Then, we prove that $F_U^*$ should have a single amplitude level. Finally, we identify the structure of the optimal input by establishing its discreteness and locating its mass points.

\subsection{Optimality of $\frac{2\pi}{2^b}$-symmetric input distribution}
In this subsection, we show that the optimal input distribution is $\frac{2\pi}{2^b}$-symmetric (i.e. $U \sim e^{j\frac{2\pi k}{2^b}}U$ for all $k\in\mathbb{Z}$). We first prove a key lemma about the properties of $W_{y}^{(b)}(\alpha,\theta)$.
 \begin{lemma}
The function $W_{y}^{(b)}(\alpha,\theta)$ (or $W_{y}^{(b)}(u)$) satisfies the following properties:
\begin{align*}
    (i) && \; W_{y}^{(b)}\left(\alpha,\theta + \frac{2\pi k}{2^b}\right) =& W_{y-k}^{(b)}\left(\alpha,\theta\right)&&,\forall k\in\mathbb{Z}\\
    (ii) && \;\;W_{2^{b-1}-y}^{(b)}\left(\alpha,\frac{\pi}{2^b}\right) =& W_{2^{b-1}+y}^{(b)}\left(\alpha,\frac{\pi}{2^b}\right)  \\
    (iii) && \;W_{2^{b-1}-y}^{(b)}\left(\alpha,0\right) =& W_{2^{b-1}-1+y}^{(b)}\left(\alpha,0\right)\\
\end{align*}
\end{lemma}
\begin{proof}
See \cite[Section A]{ISIT2021_supplement}.
\end{proof}
Lemma 1.i states that shifting the input by $\frac{2\pi k}{2^b}$ corresponds to a shift in the phase quantizer output by $-k$. Meanwhile, Lemma 1.ii and 1.iii identify some symmetry of $W_{y}^{(b)}(\alpha,\theta)$ when $\theta = 0$ and $\theta = \frac{\pi}{2^b}$. The following proposition shows that the capacity is achieved by a $\frac{2\pi}{2^b}$-symmetric distribution. Thus, without loss of generality, we can simply restrict our search of $F_U^*$ in this set of input distributions.
\begin{proposition}
For any input distribution $F_U$, we define another input distribution as
\begin{equation}
    F_U^{s} = \frac{1}{2^b}\sum_{i = 0}^{2^b-1}F_U(ue^{j\frac{2\pi i}{2^b}}),
\end{equation}
which is a $\frac{2\pi}{2^b}$-symmetric distribution. Then, $I(F_U^{s}) \geq I(F_U)$. Under this input distribution, $H(Y)$ is maximized and is equal to $b$.
\end{proposition}
\begin{proof}
See \cite[Section B]{ISIT2021_supplement}.
\end{proof}
Because of Proposition 1, we consider $F_U\in \Omega_{s}$, where $\Omega_{s}$ is the set of all $\frac{2\pi}{2^b}$-symmetric input distributions satisfying the constraint $\mathbb{E}[|U|^2] \leq P'$. The capacity in (\ref{eq:cap_def}) simplifies to
\begin{equation}\label{eq:cap_sym}
    \begin{split}
        C = b - \underset{F_U\in\Omega_s}{\inf} \int_{\mathbb{C}}\underbrace{-\sum_{y=0}^{2^b-1}W_{y}^{(b)}(u)\log W_{y}^{(b)}(u)}_{H(Y|U=\sqrt{\alpha}e^{j\theta})}\;dF_U.
    \end{split}
\end{equation}

\subsection{Optimality of input with a single amplitude level}
To prove that the optimal input should have a single amplitude level, we first establish two properties of $H(Y|U=u)$.
\begin{lemma}
The function $H(Y|U=\sqrt{\alpha}e^{j\theta})$ is decreasing on $\alpha$ for all $\theta \in \left[0,\frac{2\pi}{2^b}\right)\text{ and } b\geq 1$.
\end{lemma}
\begin{proof}
See \cite[Section C]{ISIT2021_supplement}.
\end{proof}

\begin{lemma}
The function $H(Y|U=\sqrt{\alpha}e^{j\theta})$ is convex on $\alpha$ for all $\theta \in \left[0,\frac{2\pi}{2^b}\right)\text{ and } b\geq 1$.
\end{lemma}
\begin{proof}
See \cite[Section D]{ISIT2021_supplement}.
\end{proof}

The capacity in (\ref{eq:cap_sym}) can be written as
\begin{equation*}
    \begin{split}
        C =& b - \underset{F_U\in\Omega_s}{\inf} \mathbb{E}_{A,\Theta}\left[-\sum_{y=0}^{2^b-1}W_{y}^{(b)}(u)\log W_{y}^{(b)}(u)\right]\\
        =& b - \underset{F_U\in\Omega_s}{\inf} \mathbb{E}_{\Theta}\left[\mathbb{E}_{A|\Theta}\left[-\sum_{y=0}^{2^b-1}W_{y}^{(b)}(u)\log W_{y}^{(b)}(u)\right]\right],
    \end{split}
\end{equation*}
where we used Bayes' rule in the second line to express the complex PDF $f_U(u) = f_{A,\Theta}(\alpha,\theta)$ as $f_{A|\Theta}(\alpha|\theta)f_{\Theta}(\theta)$ and perform the complex expectation as two real-valued expectations over $\alpha|\theta$ and $\theta$. Due to Lemma 3, Jensen's inequality can be applied. That is,
\begin{align*}
    \mathbb{E}_{A|\Theta}\left[H(Y|U=\sqrt{\alpha}e^{j\theta})\right] \geq H(Y|U=\sqrt{\mathbb{E}_{\alpha|\Theta}[\alpha]}e^{j\theta}),
\end{align*}
with equality if $\alpha$ is a constant. This means that for some $\frac{2\pi}{2^b}$-symmetric input distribution, $F_{U}^{(a)}$, with two or more amplitude levels, there exists another $\frac{2\pi}{2^b}$-symmetric input distribution, $F_{U}^{(b)}$, with one amplitude level that has lower $\mathbb{E}_{A|\Theta}\left[H(Y|U=\sqrt{\alpha}e^{j\theta})\right]$ than $F_{U}^{(a)}$. Moreover, due to Lemma 2, for any $\frac{2\pi}{2^b}$-symmetric input distribution with amplitude $\alpha_a < P'$, we can find another $\frac{2\pi}{2^b}$-symmetric input distribution with amplitude $\alpha_b \in (\alpha_a,P']$ such that $\mathbb{E}_{A|\Theta}\left[H(Y|U=\sqrt{\alpha_a}e^{j\theta})\right] > \mathbb{E}_{A|\Theta}\left[H(Y|U=\sqrt{\alpha_b}e^{j\theta})\right]$. Thus, full transmit power must be used. We formalize this result in the following proposition.
\begin{proposition}\label{proposition:constant_amplitude}
The optimum input distribution has a single amplitude level $\sqrt{\alpha} = \sqrt{P'}$. 
\end{proposition}
The capacity expression can be simplified further to
\begingroup
\allowdisplaybreaks
\begin{align*}
    C =& b - \underset{F_\Theta\in\Omega_{\Theta}^{s}}{\inf} \mathbb{E}_{\Theta}\left[-\sum_{y=0}^{2^b-1}W_{y}^{(b)}(P',\theta)\log W_{y}^{(b)}(P',\theta)\right],
\end{align*}
\endgroup
where $\Omega_{\Theta}^{s}$ is the set of all circular distributions with support $[-\pi,+\pi]$ that are $\frac{2\pi}{2^b}$-symmetric. That is, 
\[F_{\Theta}(\theta) \sim F_{\Theta}\left( \left(\theta + \frac{2\pi k}{2^b}\right)\mod 2\pi \right),\quad\forall k\in\mathbb{Z}.\]
\subsection{Discreteness of the Optimal Input and Location of its Mass Points}
We continue with the derivation of the optimum input distribution by identifying the minimizer of the optimization problem
\begin{equation}\label{eq:optprob_phase_quant_entrop}
    \underset{F_\Theta\in\Omega_{\Theta}^{s}}{\inf} \mathbb{E}_{\Theta}\left[-\sum_{y=0}^{2^b-1}W_{y}^{(b)}(P',\theta)\log W_{y}^{(b)}(P',\theta)\right].
\end{equation}
We present two lemmas about the objective function and feasible set of (\ref{eq:optprob_phase_quant_entrop}).
\begin{lemma} The set $\Omega_{\Theta}^s$ is convex and weakly compact.
\end{lemma}
\begin{proof}
See \cite[Section E]{ISIT2021_supplement}.
\end{proof}
\begin{lemma} The function 
\begin{equation}
    \bar{w}(F_{\Theta}) = \mathbb{E}_{\Theta}\left[-\sum_{y=0}^{2^b-1}W_{y}^{(b)}(P',\theta)\log W_{y}^{(b)}(P',\theta)\right]
\end{equation}
is convex and weakly differentiable on $F_{\Theta}$.
\end{lemma}
\begin{proof}
See \cite[Section F]{ISIT2021_supplement}.
\end{proof}
The combination of Lemma 4 and Lemma 5 implies that Problem (\ref{eq:optprob_phase_quant_entrop}) is a convex optimization problem over the probability space $\Omega_{\Theta}^s$. An optimal solution $F_{\Theta}^*$ should satisfy the following inequality:
\[ \bar{w}'_{F_{\Theta}^*}(F_{\Theta}) = \bar{w}\left(F_{\Theta}\right)-\bar{w}\left(F_{\Theta}^*\right) \geq 0\qquad\forall F_{\Theta}\in\Omega_{\Theta}^s,
\]
where $\bar{w}'_{F_{\Theta}^0}(F_{\Theta})$ is the weak derivative\footnote{The notions of weak derivative and weakly differentiable functions are introduced in \cite[Section F]{ISIT2021_supplement}.} of $\bar{w}(F_{\Theta})$ at a point $F_{\Theta}^0$. With some manipulation, the optimality condition can be established as
\begingroup
\allowdisplaybreaks
\begin{align*}
    \bar{w}\left(F_{\Theta}\right)-\bar{w}\left(F_{\Theta}^*\right) \geq& 0\\
    b-\bar{w}\left(F_{\Theta}^*\right) -b+\bar{w}\left(F_{\Theta}\right) \geq& 0\\
    C -b+\mathbb{E}_{\Theta}\left[-\sum_{y=0}^{2^b-1}W_{y}^{(b)}(P',\theta)\log W_{y}^{(b)}(P',\theta)\right] \geq& 0,
\end{align*}
where the third line follows from the definition of capacity. Finally, by applying the contradiction argument in \cite[Theorem 4]{Abou-Faycal:2001}, we obtain
\begin{align}\label{eq:KTC}
    C -b-\sum_{y=0}^{2^b-1}W_{y}^{(b)}(P',\theta)\log W_{y}^{(b)}(P',\theta) \geq& 0,
\end{align}
\endgroup
with equality if $\theta\in F_{\Theta}^*$. We further reduce the search space by proving that the optimal input distribution is discrete with finite number of mass points. The proof closely follows the example application of Dubin's Theorem \cite{Dubins:1962} presented in \cite[Section II-C]{Witsenhausen:1980}. 
\begin{lemma}
The support set of $F_{\Theta}^*$ is discrete and contains at most $2^b$ points.
\end{lemma}
\begin{proof}
See \cite[Section G]{ISIT2021_supplement}.
\end{proof}
Due to Proposition 1, we can limit our search of $\theta^*$ in $[0,\frac{2\pi}{2^b})$ since if $\theta^* \in [0,\frac{2\pi}{2^b})$ is optimal, so are $\theta^* + \frac{2\pi k }{2^b}$ for $k\in\mathbb{Z}$. Moreover, the optimal distribution has a single mass point inside $[0,\frac{2\pi}{2^b})$ as a consequence of Lemma 6 and Proposition 1. Because the only way to place a nonzero number of mass points in a $\frac{2\pi}{2^b}$-symmetric input distribution that is less than or equal to $2^b$ is to have exactly one mass point at every $\theta_0 + \frac{2\pi k}{2^b}$ for $k\in\mathbb{Z}$ and for some $\theta_0 \in \left[0,\frac{2\pi}{2^b}\right)$. A $\frac{2\pi}{2^b}$-symmetric distribution cannot be achieved by using less than $2^b$ mass points. Moreover, these mass points should have equal amplitudes and are equiprobable to satisfy Proposition 1. We now utilize (\ref{eq:KTC}) in Proposition 3 to obtain the location of the optimal mass points.

\begin{proposition} The set containing the angles of the optimum mass points $u^*\in F_U^*$ is given by
\begin{equation}
    \theta^* = \left\{\frac{2\pi(k+0.5)}{2^b}\right\}_{k = 0}^{2^b-1}.
\end{equation}
\end{proposition}
\begin{proof}
See \cite[Section H]{ISIT2021_supplement}.
\end{proof}

Simply put, Proposition 3 states that the optimal location of the mass points are at the angle bisector of the convex cones $\mathcal{R}_{y}^{\text{PH}}$. Now that we have established the characteristics of the $F_U^*$, we formally state in the following theorem the capacity of the system.

\begin{figure}[t]
\centering
\hspace*{-.3 cm}
    \includegraphics[scale = .7]{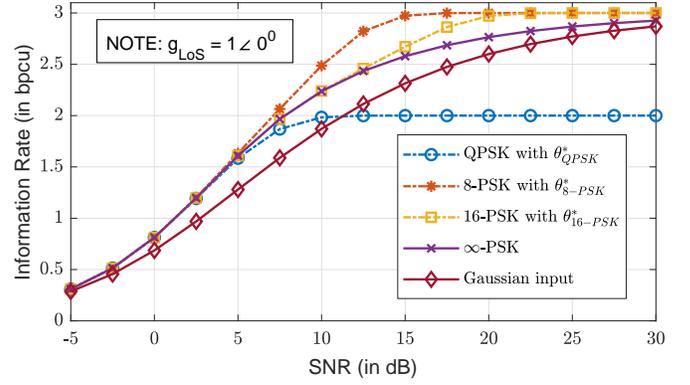}
    \caption{Information rates achieved by different modulation schemes when $g_{\text{LoS}} = 1 \angle 0^0$ and $b = 3$. Note that 8-PSK with optimal $\theta$ is capacity-achieving.}
    \label{fig:rate_vs_snr_AWGN}
\end{figure}

\begin{theorem}\label{theorem:AWGN_case} The capacity of a complex Gaussian channel with fixed channel gain and $b$-bit phase-quantized output is
\begingroup
\allowdisplaybreaks
\begin{align}\label{eq:cap_AWGN}
    C = b + \sum_{y=0}^{2^b-1}W_{y}^{(b)}\left(P',\frac{\pi}{2^b}\right)\log W_{y}^{(b)}\left(P',\frac{\pi}{2^b}\right),
\end{align}
\endgroup
and the capacity-achieving input distribution is a rotated $2^b$-PSK with equiprobable symbols given by
\begin{equation*}
    f_X^* = \left\{\frac{\delta(x)}{2^b}\Big|x = \sqrt{P}e^{j\left(\frac{2\pi(k+0.5)}{2^b}-\angle g_{\text{LoS}}\right)}, k\in[0,2^{b}-1]\right\}.
\end{equation*}
\end{theorem}
\begin{proof}
The proof follows from calculating (\ref{eq:cap_def}) using $F_U^*$. The capacity-achieving input $F_X^*$ follows from combining Propositions 1-3 and using the transformation $X = \sigma U/g_{\text{LoS}}$. 
\end{proof}
To demonstrate the optimality of the signaling scheme, Figure \ref{fig:rate_vs_snr_AWGN} compares the rates achieved by using 4,8,16, and $\infty$-PSK (a circle) with equiprobable mass points on a Gaussian channel with 3-bit phase-quantized output. Each PSK constellation is rotated by a $\theta^*$ that maximizes the rate. The rate of Gaussian input is also included and is seen to be suboptimal compared to $\frac{\pi}{4}$-symmetric input distributions with a single amplitude. It can be observed that 8-PSK with optimal $\theta$ achieves the highest rate among all modulation orders considered.

\section{Outage Probability of Rayleigh Channel with Phase-Quantized Output}\label{section-outage-prob}

\begin{figure}[t]
    \hspace{0cm}
    \includegraphics[scale = .7]{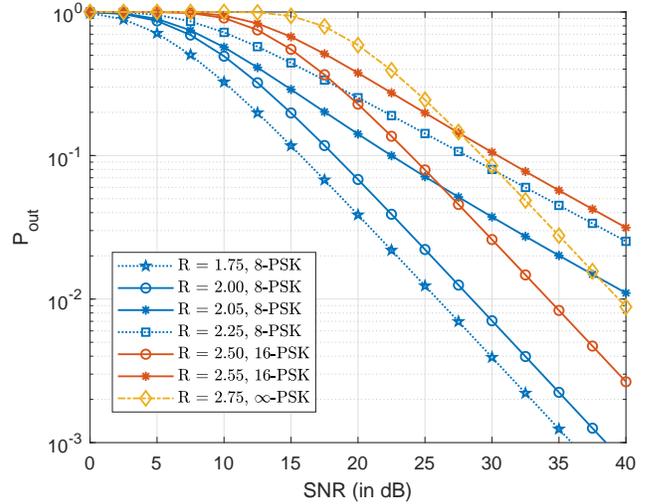}
    \caption{Outage Probability $P_{\text{out}}$ vs. SNR for different rate $R$ and PSK modulation $\mathcal{X}_M$ ($b = 3$).}
    \label{fig:outage_prob}
\end{figure}

\begin{figure}[t]
    \hspace{-.5cm}
    \includegraphics[scale = .675]{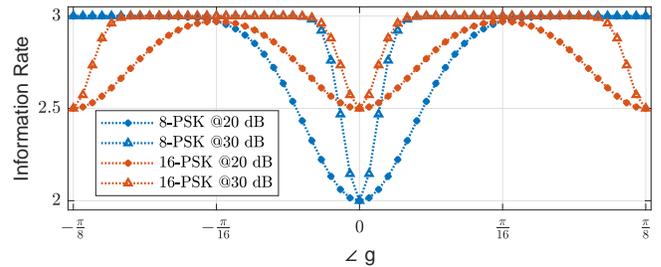}
    \caption{Rate vs. $\angle g$ for $\mathcal{X}_M = \{\text{8-PSK,16-PSK}\}$ ($|g| = 1$).}
    \label{fig:rate_vs_angle}
\end{figure}

We have shown that $2^b$-PSK is optimal for an AWGN channel with $b$-bit phase-quantized output and $h$ is known at the transmitter. We now ask if this continues to hold when channel information is unavailable at the transmitter. Ultimately, is $2^b$-PSK still the best choice in fading environment without channel state feedback? We now consider a quasi-static Rayleigh flat fading environment. The fixed channel gain $g_{\text{LoS}}$ in Figure \ref{fig:psk_receiver_sys_model} is replaced by a random fading gain $G\sim \mathcal{CN}(0,1)$. We further assume that the fading state $g$ is known only at the receiver. Without loss of generality, we assume $\sigma^2 = 1$. We define the function
\begin{equation*}
    \begin{split}
        I(\mathcal{X}_M|G=g) =& b -  \mathbb{E}_{B}\left[H\left(Y|U = \sqrt{|g|^2SNR}e^{j(\beta+\angle g)}\right)\right]\\
        =&b - r\left(|g|^2SNR,\angle g,\mathcal{X}_M\right)
    \end{split}
\end{equation*}
as the maximum rate of reliable communication supported by a modulation scheme $\mathcal{X}_M$ and a fading realization $g$ at some SNR. Here, the symbols $x\in \mathcal{X}_M$ have the form $x = \sqrt{SNR}e^{j\beta}$ so that Propositions 1 and 2 are satisfied\footnote{We omit the proof that these necessary conditions for optimum $\mathcal{X}_M$ hold even when $g$ is unknown at the transmitter.}. If the transmitter encodes the data at a rate $R$ bits/channel use, then an outage occurs when $I(\mathcal{X}_M|G=g) < R$ since the error rate cannot be made arbitrarily small whatever coding scheme is used. The function $r\left(\gamma,\angle g,\mathcal{X}_M\right)$ is a convex decreasing function of $\gamma$ (Lemmas 2 and 3). Thus, it follows that its inverse function with respect to $\gamma$ has one-to-one mapping and is also decreasing. The outage probability is expressed as
\begin{align}
        P_{\text{out}}(SNR) =& \mathbb{E}_{G}\left[\mathbb{P}\left\{I(\mathcal{X}_M|G=g)< R\right\}\right]\nonumber\\
        =&\mathbb{E}_{G}\left[\mathbb{P}\left\{\frac{r^{-1}(b-
        R,\angle g,\mathcal{X}_M)}{SNR} <  |g|^2\right\}\right]\nonumber\\
        =& 1 - \int_{-\pi}^{\pi}\;\frac{\exp\left(-\frac{r^{-1}(b-R,\angle g,\mathcal{X}_M)}{SNR}\right)}{2\pi}\text{d}\angle g.
\end{align}
The third line follows by noting that $|g|^2$ is exponentially-distributed for Rayleigh fading and $\angle g$ is uniformly-distributed. It is difficult to analytically derive the outage probability so the expression for $P_{\text{out}}(SNR)$ is evaluated numerically to provide some more insight. In order to characterize the outage probability, we focus on the outage exponent (or diversity order) which is the asymptotic slope of the outage probability as a function of SNR. Mathematically, this is defined as 
\begin{align}
    \text{DVO} = \underset{SNR\rightarrow \infty}{\lim} -\frac{\log P_{\text{out}}(SNR)}{\log SNR}.
\end{align}
Figure \ref{fig:outage_prob} depicts the outage probability of Rayleigh fading channel with 3-bit phase-quantized output for different $R$ and $\mathcal{X}_M = \{\text{8-PSK, 16-PSK, $\infty$-PSK}\}$. One noteworthy observation is the sudden decrease of the outage exponent when $R$ is increased from $2.00$ to $2.05$ for 8-PSK. DVO drops from $1$ to $\frac{1}{2}$. This is also the case for 16-PSK when $R$ is increased from $2.50$ to $2.55$. This can be partially explained by rates of 8-PSK and 16-PSK for varying rotations (as seen in Figure \ref{fig:rate_vs_angle}). Since the transmitter cannot compensate the phase rotation induced by fading, choosing an $R$ that exceeds the worst-case rates of 8-PSK and 16-PSK in Figure \ref{fig:rate_vs_angle} causes outage even with high SNR. Lastly, we note that $\infty$-PSK is invariant of the channel phase. As such, the choice of $R$ does not affect its outage exponent provided $R < b$. However, the input distribution that achieves the best outage performance for a particular SNR and quantizer resolution still needs to be addressed by further research.

\section{Conclusion}

In this work, we analyzed the capacity of channels with phase-quantized output. The first contribution of this work is a rigorous proof that a rotated $2^b$-phase shift keying is optimal for static channels with $b$-bit phase quantization. Using the capacity-achieving input, a channel capacity expression is established. Numerical examples were provided to demonstrate the optimality of the capacity-achieving input. For phase-quantized Rayleigh fading case, the outage performance was analyzed numerically for different $M$-PSK modulation schemes and $b = 3$. Our numerical findings showed that transmitting at a rate $R$ that is above the information rate of $M$-PSK signaling with worst-case $\angle g + \beta$ would significantly impact the robustness of the system against outage. A threshold effect in the outage exponent was observed in 8-PSK and 16-PSK when $R$ exceeded these values. Further research needs to be conducted to be able to generalize these results to different types of fading channels. Ergodic capacity of phase-quantized fading channel is also considered for future work.




\ifCLASSOPTIONcaptionsoff
  \newpage
\fi

\nocite{Alaoglu:1940}
\bibliographystyle{ieeetr}
\bibliography{references}

\end{document}